# Broadening the Bandwidth of Metamaterial Cloaks with Non-Foster Metasurfaces

Pai-Yen Chen, Christos Argyropoulos, and Andrea Alù

The University of Texas at Austin, Department of Electrical and Computer Engineering, Austin, TX, 78712, U.S.A.

*We introduce the concept and practical design of broadband, ultrathin cloaks based on non-Foster, negatively capacitive metasurfaces. By using properly tailored, active frequency-selective screens conformal to an object, within the realm of practical realization, is shown to enable drastically reduced scattering over a wide frequency range in the microwave regime, orders of magnitude broader than any available passive cloaking technology. The proposed active cloak may impact not only invisibility and camouflaging, but also practical antenna and sensing applications.*

Metamaterials and plasmonic materials have attracted wide attention in recent years due to their exciting applications, which include the possibility of suppressing the visibility of an object by realizing suitably designed cloaking layers [1]-[10]. Beyond camouflaging, metamaterial cloaks have been proposed for non-invasive probing [11], high-fidelity biomedical measurements [12], low-noise communications [13], optical tagging [14], nonlinear nanodevices [15] and scattering signature manipulation [16]. The drastic suppression of an object's scattering is generally possible based on the anomalous wave interaction with artificially engineered metamaterials, provided that their material dispersion and spatial profiles meet the required theoretical requirements. Despite recent advances in metamaterial engineering [17], so far the practical realization of these devices has faced significant challenges, due to the stringent requirements on material properties and fundamental constraints on the overall bandwidth of operation. This problem is particularly evident for transformation-based cloaks [3]-[8], which inherently require extreme material properties, anisotropy and inhomogeneity profiles. To meet these requirements, resonant and dispersive inclusions are necessary, implying severe bandwidth limitations and high sensitivity to disorder and imperfections. Solutions to relax these limitations have been suggested by reducing the requirement on maximum scattering suppression [18], but inherent limitations associated with causality and passivity appear to fundamentally limit the cloaking performance of any available technology that aims at suppressing the total scattering from an object in all directions and for all excitations [19]-[23].

Recently, we proposed and experimentally verified an alternative cloaking technique, named *mantle cloaking* [24]-[27], based on the concept of "cloaking by a surface", for which a suitably designed metasurface supports current distributions radiating 'anti-phase' fields that cancel the scattering from the covered object. Mantle cloaks can be readily realized at microwaves by patterning a metallic surface around the object of interest, and various structural designs [see Fig. 1(a)] have been proposed in the context of metasurfaces and frequency-selective surfaces (FSS) [28]. We have recently shown that even a one-atom-thick graphene monolayer may achieve scattering suppression at THz frequencies [29]. The ultrathin profile of mantle cloaks makes their practical realization easier than bulk metamaterial cloaks, and it is also usually associated with a moderate bandwidth improvement compared with the other cloaking techniques based on bulk metamaterials [24]. Still, similar bandwidth limitations apply to this scenario. Achieving broadband cloaks, on the contrary, may be disruptive for several of the above mentioned applications, and may finally bring metamaterials close to a practical impact on technology.

In order to understand the fundamental limitations on bandwidth specifically applied to the mantle cloaking technique, consider the simple case of an infinite cylinder with relative permittivity $\varepsilon = 3$, relative permeability $\mu = 1$ and diameter $2a = 12 \, \text{cm}$, covered by a two-dimensional conformal mantle cloak with radius $a_c$ [see Fig.1(c)]. If the periodicity of the metallic patterns is smaller than the wavelength of operation, under certain symmetries of the excitation we can describe the metasurface with an isotropic averaged surface impedance $Z_s = R_s - iX_s$ under an $e^{-i\omega t}$ time convention [24],[26].

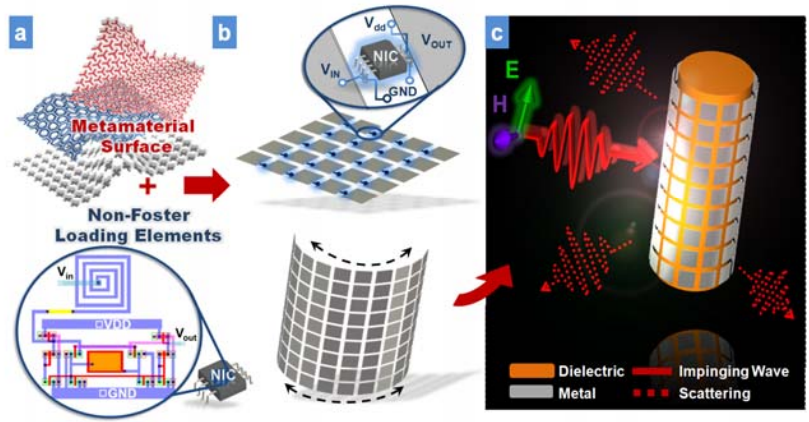

*Figure 1* Schematics of (a) an ultrathin metasurface formed by structured metal (top) combined with lumped NICs based on CMOS technology; (b) assembly of an active non-Foster metasurface; (c) a mantle cloak designed for a dielectric infinite cylinder under TM illumination.

For a normally incident, transverse-magnetic (TM) plane wave with magnetic field polarized orthogonal to the cylinder axis [Fig. 1(c)], for which largest interaction is expected in the case of an object of moderate thickness, the Mie scattering coefficients may be written [24] as $c_n^{TM} = -U_n^{TM} / \left( U_n^{TM} + iV_n^{TM} \right)$, and the total scattering width (SW) of the cylinder normalized to the wavelength, a quantitative measure of its overall visibility, becomes [30] $\sigma_{2D} = \frac{2}{\pi} \sum_{n=0}^{n=\infty} \left( 2 - \delta_{n,0} \right) \left| c_n^{TM} \right|^2$. In the long-wavelength limit, these expressions simplify, as the $n = 0$ harmonic dominates the overall scattering, and the condition to achieve identically zero $c_0^{TM}$ becomes [25]:

$$X_s = \frac{2}{\omega a \gamma \varepsilon_0 (\varepsilon - 1)}, \tag{1}$$

where $\gamma = a/a_c$. This value is inductive for dielectric objects $\varepsilon > 1$, as expected due to the capacitive nature of a moderate size dielectric cylinder. However, the frequency dispersion of the required inductance violates Foster's reactance theorem, which states that the reactance of any passive element in regions of low absorption must monotonically increase with frequency, $\partial X_s(\omega)/\partial \omega > 0$ [31]. Eq. (1) actually corresponds to a negative capacitance $Z_s = 1/\left( -i\omega \left( -C_{eff} \right) \right)$, with $C_{eff} = \varepsilon_0 (\varepsilon - 1) \gamma a / 2$, which may be achieved in a circuit only considering amplifiers or active elements, for which Foster's theorem does not apply. Essentially the metasurface is required to catch up with the capacitive dispersion of the object, requiring an active loading. This is consistent with the idea of relaxing the bandwidth limitations of bulk metamaterials with the use of active inclusions [32]. Indeed, concepts to employ non-Foster elements and active metamaterials to realize broader bandwidths in a variety of applications are being currently explored [33]-[37].

Even if we may intuitively expect that active inclusions may overcome the passivity constraints mentioned above, in practice it is challenging to design a practical active metamaterial cloak realizing this effect. In the following, on the contrary, we show that a non-Foster mantle cloak may be realistically formed by loading a subwavelength metallic patch array with negative impedance converter (NIC) elements. The mantle cloaking technique is particularly well suited to be combined with lumped NIC elements, and allows combining a large bandwidth of operation with ultra-low profile and relatively simple realization.

Moving to the full-wave dynamic scenario, Fig. 2(a) shows the optimal surface reactance $X_{s,opt}$ (red solid line) required to minimize the total scattering from the dielectric cylinder of Fig. 1 at every frequency point using a lossless mantle cloak with radius $a_c = a$. The curve was calculated using the rigorous analytical formulation developed in [25]: it is obvious that a non-Foster dispersion is indeed required to realize broadband cloaking, as $\partial X_{s,opt}(\omega)/\partial \omega < 0$ over the whole frequency range of interest. First, let us assume to use a passive mantle cloak with inductive reactance $jX_{s,MTS} = j\omega L$ $[\Omega]$ and $L = 62.620$ nH, designed to suppress the scattering at the design frequency $f_0 = 0.8$ GHz. Its frequency dispersion (blue solid line) can intersect the optimal reactance curve $X_{s,opt}$ at one single point $X_{s,FSS}\big|_{f_0} = 314.76\,\Omega$, implying that the achievable cloaking bandwidth is intrinsically narrow, as dictated by causality and passivity. Even using a multilayered cloak [2], we may be able to hit the optimal curve at various frequencies, but Foster's theorem would require a scattering peak to arise in between every two zeros [14],[38], inherently limiting the available cloaking bandwidth. Figure 2(b) shows the SW frequency variation for the passive cloaked cylinder (blue solid line), compared to the uncloaked case (gray dashed line). Indeed, the passive cloak can significantly suppress the scattering around the design frequency $f_0$, but only over a limited bandwidth. Moreover, at lower frequencies the cloaked cylinder generates more scattering than the uncloaked case, due to its inherent frequency dispersion.

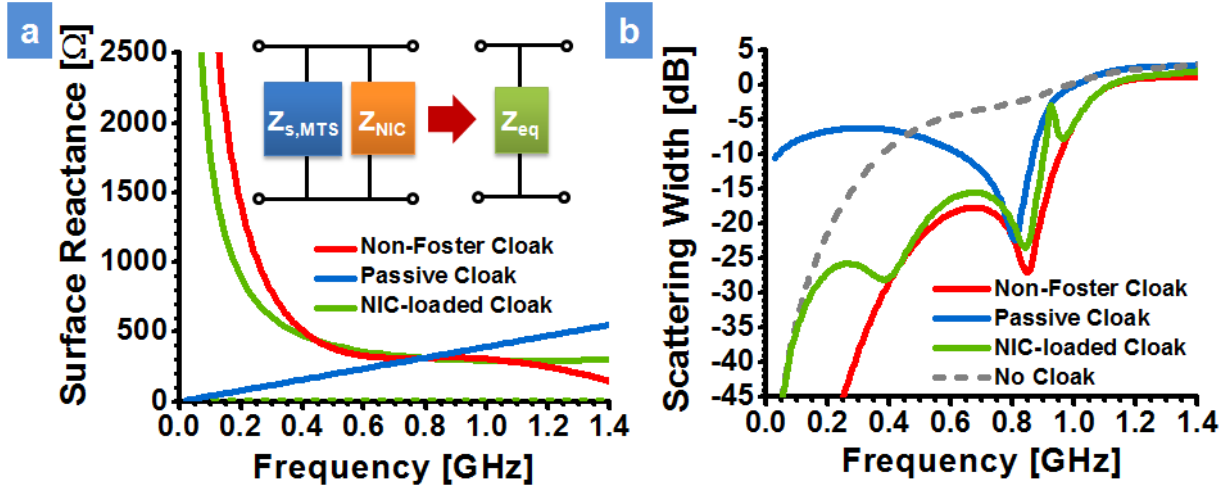

*Figure 2 - (a) Variation of the surface reactance for: an optimal (non-Foster) mantle cloak (red), a passive mantle cloak designed to cloak at the design frequency $f_0$ (blue), a realistic NIC-loaded (green) mantle cloak; the green dashed line shows the corresponding surface resistance. (b) Variation of the normalized SW for the dielectric cylinder covered by the cloaks in (a). The uncloaked scenario (gray dashed) is also shown for comparison.*

To realize a broadband cloak, we must rely on active metasurfaces that can break Foster's limitations. Our design starts from a passive metasurface made of an array of metal square-patches [Fig. 1(b)], with an array period $d=1.8$ cm and a gap between neighboring elements $g = 0.12\, d$. This geometry is well described by the equivalent capacitive impedance $Z_{patch}^{TM} = R_{patch} + 1/(j\omega C_{patch})$; detailed expressions for $R_{patch}$ and $C_{patch}$ in terms of the geometry parameters, and their applicability to mantle cloaks, are available in [26],[39]. When the gaps are loaded by lumped circuit elements $Z_{NIC}$, as in Fig. 1a, the equivalent surface impedance is given by the parallel combination of the load impedance and the metasurface impedance [inset of Fig.

2(a)] $Z_{s,eq} = R_{s,eq} + jX_{s,eq} = \left(\left(Z_{patch}^{TM}\right)^{-1} + \left(Z_{NIC}\right)^{-1}\right)^{-1}$ [40]. We consider here active NIC loads connecting neighboring patches, breaking the non-Foster constraints on the surface impedance. The simplest NIC design with such desired functionality is based on a cross-coupled pair of transistors [33] (i.e., bipolar junction transistors). However, to be able to tailor $Z_{s,eq}$ to fit the optimal frequency dispersion $X_{s,opt}$ over a broad range of frequencies, and at the same time ensuring that the underlying circuit is stable and practically feasible, we propose a more complex active circuit design, including a network of passive, resistive and reactive components. Details of our design layout based on microelectronic and silicon-chip technology (TSMC 0.18 $\mu$m process) are presented in [39].

Figure 2(a) shows the calculated surface reactance (green solid line) of our final design, including all parasitic effects expected in the realization of this metasurface. It is seen that the electromagnetic response can be well tailored to follow $X_{s,opt}$ (red solid line) over a broad frequency range from 0.4 GHz to 1.2 GHz (with a relative error less than 5 %). Since our design requirement is to keep the SW below -20 dB, the optimized reactance does not follow $X_{s,opt}$ at very low frequencies, at which the uncloaked cylinder anyway scatters very little. Figure 2(a) also presents the surface resistance of this active metasurface (green dashed line), verifying that the loss of the proposed non-Foster mantle cloak is small yet positive across all the considered frequency range. This ensures that the scattering response is stable, despite the presence of active elements, and that at the same time the cloak is reasonably low-loss. In this regard, we stress that we are not aiming here at broadening a resonant response, which may inherently cause stability issues as in recent works on non-Foster impedance matching [41], but instead we suppress the overall scattering of a passive object, which is an inherently nonresonant process, more prone to remain stable. We prove in [39] that the overall system, active metasurface and passive cylinder, is unconditionally stable. Figure 2(b) shows the corresponding SW for the realistic NIC-loaded mantle cloak (green solid line). The cloak provides a drastically improved bandwidth, much broader than an ideal passive cloak, with a normalized SW well suppressed below -15 dB up to approximately 900 MHz. For very low frequencies the NIC-loaded mantle cloak does induce slightly more scattering than an ideal non-Foster cloak, but in this region the object itself has a very low scattering signature because of its small electrical size.

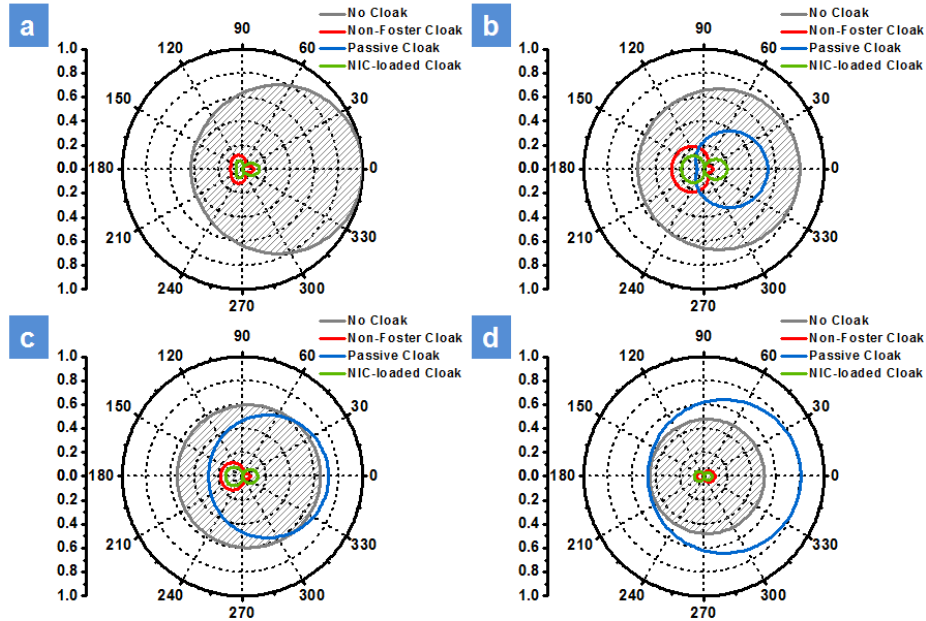

*Figure 3 -* *(a)-(d) Far-field scattering pattern of the infinite dielectric cylinder without cover (gray shadow), with the ideal non-Foster mantle cloak (red line), with the NIC-loaded mantle cloak (green line), and with the ideal passive cloak (blue line), for different frequencies of operation: (a) 0.8 GHz, (b) 0.65 GHz, (c) 0.5 GHz, and (d) 0.4 GHz.*

Figure 3 shows the calculated far-field scattering patterns for the dielectric cylinder covered by the ideal non-Foster mantle cloak (red line), our NIC-loaded mantle cloak (green line), and the inductive cloak (blue line), compared to the uncloaked cylinder (gray shadow) at various frequencies: (a) 0.8 GHz, (b) 0.65 GHz, (c) 0.5 GHz, and (d) 0.4 GHz. In all panels, the scattered fields are plotted in the same scale for fair comparison. It is evident that all cloaks provide excellent scattering suppression at the design frequency $f_0$ (panel a). However, if we observe the far-field scattering away from the design frequency, the passive cloak displays poor performance, compared to the ideal non-Foster or NIC-loaded mantle cloaks. For lower frequencies a passive cloak induces an even larger scattering than that of the uncloaked dielectric cylinder.

As the frequency increases, the performance of all cloaks deteriorates, due to the excitation of higher-order scattering harmonics, yet the active devices provide improved and more robust performance. The proposed NIC-loaded mantle cloak follows with good agreement the performance of the ideal, optimal mantle cloak, and therefore achieves the best scattering suppression achievable with a single metasurface over the whole frequency range of interest. Better results, and cloaking for larger objects, aiming at suppressing at the same time multiple scattering harmonics, may be achieved with a multilayer design. The proposed broadband cloak, whose operation covers the entire UHF band, may be of particular interest for a wide range of communication applications, beyond camouflaging. Since this technique allows the wave to enter the cloak and interact with the cloaked object, it may enable exciting applications such as broadband cloaked sensing, non-invasive probing [2] and low-interference communications [13].

We also analyze the time-domain response of the proposed mantle cloaks of Fig. 2 for a short pulse excitation in Fig. 4. This shows how the proposed active device may successfully realize a stable, invisible obstacle for short broadband pulsed excitation. Four receivers are placed in different positions along the z axis, as shown in Fig. 4(a). The distance $d_i$ between the $i$-th receiver/processor $R_{x,i}$ and the origin (center of the cylinder) is $d_1 = -16a$, $d_2 = -6a$, $d_3 = 1.5a$ and $d_4 = 6a$. Therefore, $R_1$ and $R_2$ are hit by the impinging signal before the object, whereas $R_3$ and $R_4$ are placed behind the object.

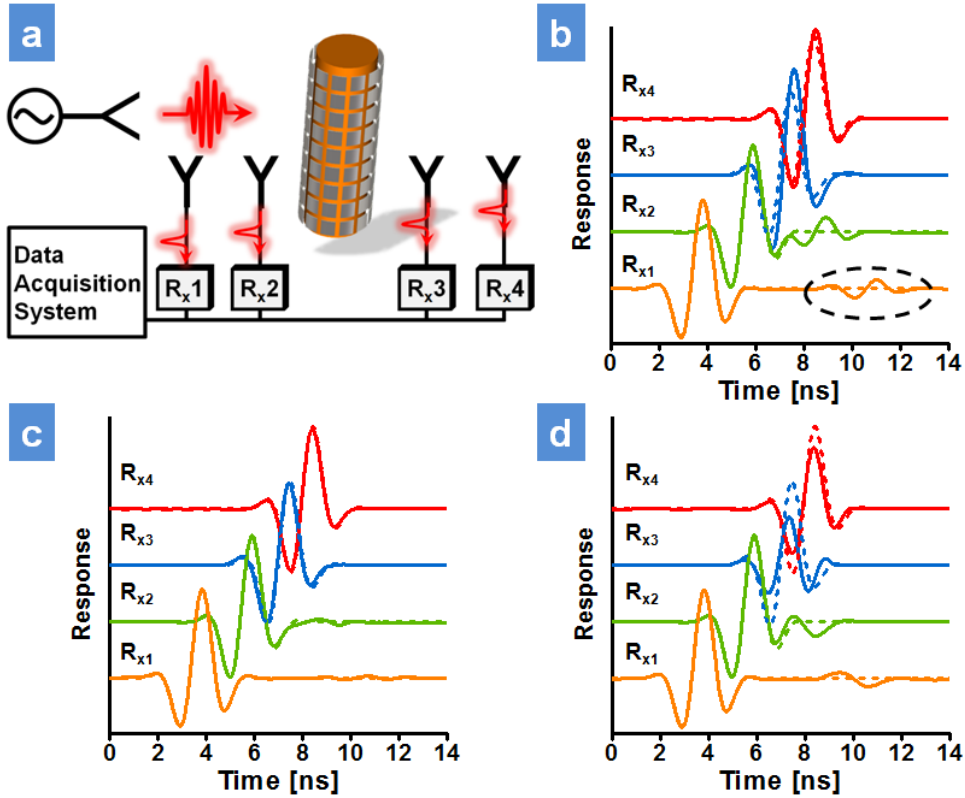

*Figure 4 - (a) Schematic diagram of the time-domain analysis of a cloaked cylinder. Signals are detected by receivers placed in different positions along the z axis, considering the cases of: (b) no cloak, (c) NIC-loaded mantle cloaks and (d) ideal passive cloak, as in Figure 2. Here the transient responses of a short Gaussian pulse traveling in free space (dashed lines) are also shown for comparison.*

Figure 4 shows the calculated transient responses at the different receivers for a short Gaussian pulse with frequency components $0.02 - 0.9\,\text{GHz}$ traveling in free-space, comparing the received signals with (solid lines) and without (dashed lines) the cylindrical scatterer. Different scattering scenarios are considered: the uncloaked cylinder [Fig. 4(b)], the NIC-loaded mantle cloak [Fig.

4(c)], and the ideal passive mantle cloak [Fig. 4(d)]. It is seen that the proposed NIC-loaded non-Foster cloak suppresses most of the signal distortion and reflections behind the object. The short pulse shape is restored to the one in absence of the cylinder, both behind and passed the object, implying that its overall bandwidth performance is excellent and stability is preserved, despite the active elements in the cloak. Both the uncloaked and the passive-cloak, on the contrary, show severely distorted signals, dispersed and delayed in time, as it is particularly obvious for signals received by $R_{x1}$ and $R_{x2}$. The passive cloak slightly improves transmission, since it cancels the scattering around $f_0$. However, the remaining frequency components contribute to distorting and stretching the tail and precursor of the signal, due to the relatively narrow cloaking bandwidth. In other words, the passive cloak can be easily detected when excited by a short pulse, while the proposed active cloak has a much more robust performance.

In conclusion, we have proposed here the concept and potential realization of a broadband mantle cloak, formed by a subwavelength metasurface loaded with optimal, active NIC elements. We have demonstrated with both time- and frequency-domain analysis that drastic scattering reduction is achievable over a broad frequency range using an inherently stable, active non-Foster mantle cloak. We envision that this low-profile and broadband cloaking technology may be applied to several applications of interest at RF and microwaves, including camouflaging and invisibility, low-invasive sensing, and low noise communications. This work was supported by the AFOSR YIP award No. FA9550-11-1-0009, the DTRA YIP award No. HDTRA1-12-1-0022 and the NSF CAREER award No. ECCS-0953311.

**Supplementary material for the paper 'Broadening the Bandwidth of Metamaterial Cloaks with Non-Foster Metasurfaces'**

I. Surface Impedance of a Mantle Cloak Made of Metallic Patch Arrays

The mantle cloak proposed in the main text is made of a periodic array of metallic patches, which can be well modeled by the equivalent capacitive impedance $Z_{FSS}^{TM} = R_{patch} + 1/(j\omega C_{patch})$, with [1]-[2]:

$$R_{patch} = \frac{d}{(d-g)\sigma}$$
$$C_{patch} = \frac{\varepsilon_{top} + \varepsilon_{bottom}}{2} \frac{2d}{c\eta_0 \pi} \ln\left[\csc\left(\frac{\pi g}{2d}\right)\right] f(\alpha)$$ (s1)

where $d = 1.8$ cm is the array period and $g = 0.12\, d$ is the gap between neighboring elements, $\sigma$ is the conductivity of metal considering the realistic Ohmic loss, $\varepsilon_{top}$ and $\varepsilon_{bottom}$ are the relative permittivities of the medium above and below the surface (here $\varepsilon_{top} = 1$ and $\varepsilon_{bottom} = 1$), $\alpha$ is the incident angle and $f_{TM}(\alpha) = 1$, $f_{TE}(\theta) = 1 - \sin^2 \alpha / (2\varepsilon_{avg})$ are angular correlation functions corresponding respectively to TM- and TE- incident polarizations.

The model (s1) is well accepted in the technical literature, and has been verified by experimental and numerical results in the microwave and THz frequency regions. Recently, we have confirmed that these formulas are well applicable to model conformal mantle cloaks based on capacitive patch arrays conformal to the object. We have approached this scattering problem using an analytical method based on Mie-Lorentz scattering theory and full-wave simulations [3]-[4]. Our results showed excellent agreement between the two methods for both plane-wave incidence [3] and line-source excitation [4], validating the use of these formulas for the purpose discussed in the main paper.

**II. Practical Realization of Non-Foster Reactive Elements and Negative Impedance Converters for Non-Foster Metasurfaces**

The concept of building a negative impedance converter inherently requires the use of active devices, in order to overcome Foster's reactance limitation. The original design for a NIC originates from Linvill's pioneering work [3], which consists of two bipolar junction transistors (BJTs) and two resistors, as sketched in Fig. S1. Interestingly, this seminal concept received significant attention only in the last decade, mainly for antenna applications, probably due to the recent advances in electronics. It has been recently applied to realize broadband cancellation of the reactive component of the antenna's input impedance [6],[7].

Figure S1 shows the circuitry diagram of a simple negative capacitance converter (NCC) for a two-port device, composed of two BJTs, a passive capacitor and passive resistors. In this simple design, the ideal input impedance, neglecting parasitic effects, is given by:

$$Z_{in,NCC} = \frac{j\, g_{m1} g_{m2} R_2 + \omega C(1+g_{m2} R_1)}{\omega C g_{m1}(1+g_{m2} R_1)}. \tag{s2}$$

where $g_{mi}$ is the transconductance of the $i$-th active device. It is seen that in the limit $g_{m1}, g_{m2} \gg C$, the NCC input impedance becomes

$$Z_{NCC} = \left(\frac{R_2}{R_1}\right)\frac{1}{i\omega(-C)}, \tag{s3}$$

which indeed provides an effective non-Foster negative capacitance matching the ideal quasi-static required surface reactance to cloak a thin dielectric cylinder [Eq. (1) in our manuscript]. It is also possible to realize more general NICs using an operation amplifier [8] and/or other field effect transistors (FETs).

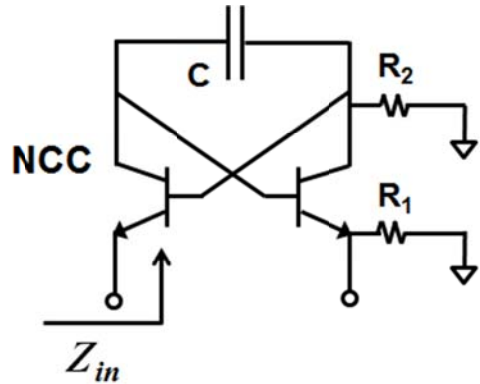

*Figure S1* - *Schematic circuit diagram for an NIC composed of a BJT pair and a capacitor connected at the output.*

An active metasurface has a surface reactance that is no longer limited by Foster's reactance theorem. Its performance and bandwidth, when used as a mantle cloak, can be significantly improved following the results in the main text, but its stability should be carefully analyzed to ensure the practical use of active elements. Unfortunately, if one attempts to combine BJTs and/or Op-Amps with rough printed circuit board (PCB) elements to realize NICs, undesired parasitic resistance in either the active devices of the NIC or the capacitor will inevitably show up at the input port as a negative resistance, causing inherent instabilities [11]. This is an especially important concern if the negative resistance is larger than the series resistance of the remaining circuit elements. In addition, for the proposed cloaking applications, the size of each lumped NIC, if realized with PCB techniques, may be comparable with the unit element of the metasurface. This makes the loading and modeling of lumped NICs very difficult. To resolve these problems, we propose to use the recent progress in radio-frequency integrated circuits

(RFIC) and complementary metal-oxide-semiconductor (CMOS) technology to realize small-sized, lumped NICs, fully integrated on-chip. This provides advantages over PCB techniques in terms of small volume, stability and predictable performance.

Figure S2(a) shows our proposed practical implementation of a NIC composed by a cross coupled metal-oxide-semiconductor field effect transistor (MOSFET) pair, a folded cascade differential amplifier to lower the DC power to $V_{DD} = 2.5\,\text{V}$, current sources and passive components including a spiral inductor, a metal resistor, and a metal-insulator-metal (MIM) capacitor. The commercial software Cadence **Error! Reference source not found.** was used to simulate and optimize this realistic circuit layout, based on TSMC's $0.18\,\mu\text{m}$ process, considering the detailed physically-based, small-signal circuit model for MOSFETs. The circuit presented in Fig. S2(a) provides enough design degrees of freedom to tailor the equivalent surface reactance to fit the non-Foster dispersion of the optimal surface reactance over the frequency range of interest. The values of passive resistors, capacitors and gate width have been fully optimized to realize a surface reactance that fits as closely as possible the optimal surface reactance dispersion in Fig. 2(a) of the manuscript, with an averaged relative error less than 5% over a broad frequency range from 0.4 GHz to 1.2 GHz. Figure S2(b) shows the final layout of the on-chip NIC circuit that may be manufactured within a compact size, much smaller than conventional NICs usually realized in PCB technology [9]-[10].

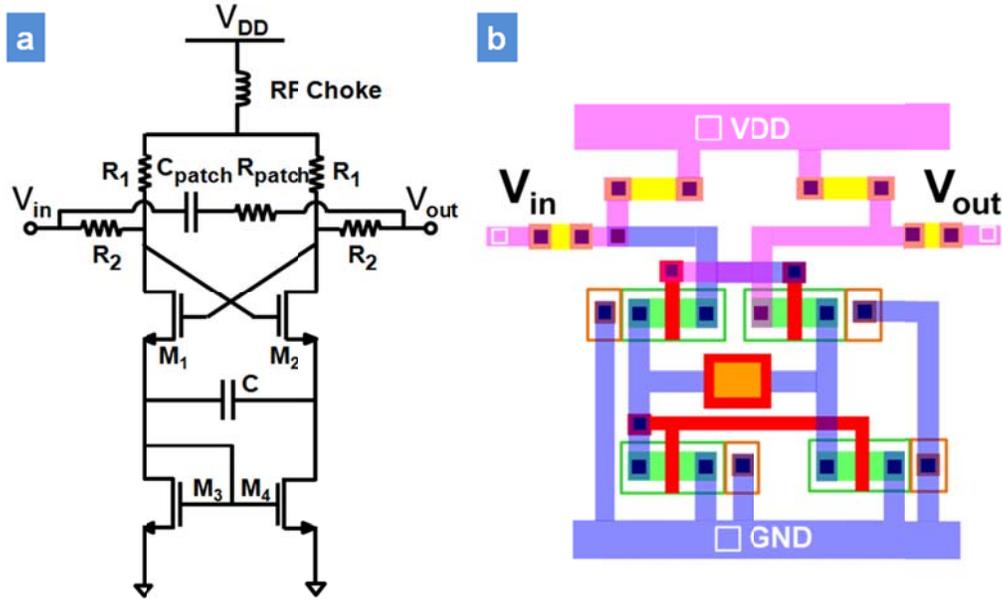

***Figure S2*** - *(a) Circuit diagram of the NIC load based on MOSFET pair and a capacitor at the output. (b) A virtual image of the layout that may be manufactured using CMOS-based microelectronic technology.*

It is imperative to examine the stability of the above circuit in order to assess its suitability for mantle cloaking. Figure S3(a) shows the reflection at the input port of the NIC (while ignoring the resistance and capacitance of the metasurface elements), showing that the reflection condition $|\Gamma| < 1$ is always satisfied over the entire spectrum of interest. This verifies that the NIC itself is stable. To achieve a more rigorous proof of stability, the unconditional stability performance should be achieved with two required conditions [11]:

(1) $\Delta < 1$, where $\Delta$ is defined as $\Delta = S_{11}S_{22} - S_{12}S_{21}$

(2) $K > 1$, where $K$ factor is defined as: $K = \dfrac{1 - |S_{11}|^2 - |S_{22}|^2 + |\Delta|^2}{2|S_{12}S_{21}|}$,

where the $S$ matrix of the two-port "black box" network is obtained with circuit simulations.

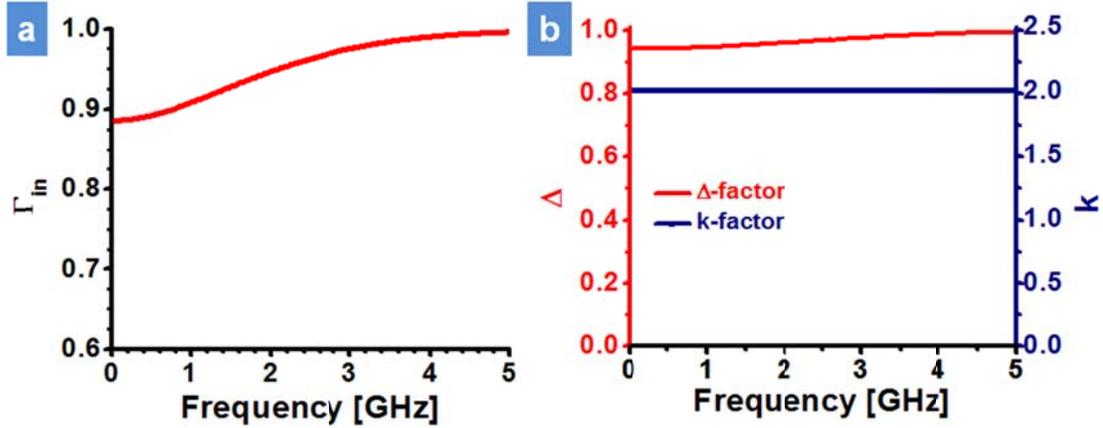

*Figure S3 - (a) Reflection at the input port of the proposed NIC device and (b) two-port stability analysis. The effect of parasitics caused by the metasurface are considered here.*

Here, we also consider the parasitic resistance and capacitance introduced by the metasurface (with relative permittivity $\varepsilon = 3$ and thickness 6 µm), which is in parallel to the NIC input impedance [Fig. S2(a)]. Figure S3(b) shows simulation results for the factors $K$ and $\Delta$. It is found that the conditions $K > 1$ and $\Delta < 1$ are satisfied over the entire spectrum, verifying unconditional stability. To achieve this operation, we need to use resistors to avoid the occurrence of a negative resistance. This, of course, introduces some small loss that partially deteriorates the cloaking performance. The equivalent surface resistance for the circuit in Fig. S2 gradually increases with frequency and becomes significantly large for frequencies above 4GHz (not shown in Fig. 2b). Typically, a compromise should be made between stability and overall cloaking performance, and for the proposed layout an optimal performance is achieved within the frequency range of interest.